\documentclass[preprint,12pt]{elsarticle}

\usepackage{amssymb}
\usepackage{amsmath}
\usepackage{graphics}
\usepackage{epsfig}

\usepackage[latin1]{inputenc}
\newdefinition{rmk}{Remark}
\newproof{pf}{Proof}

\numberwithin{equation}{section}

\usepackage{amssymb}

\journal{}

\begin{document}

\begin{frontmatter}



\title{Robust portfolio optimization using pseudodistances}

 \author[label1,label2]{Aida Toma\corref{cor1}}
\address[label1]{Department of Applied Mathematics, Bucharest Academy of Economic
Studies,\\ Pia\c{t}a Roman\u{a} 6, Bucharest, Romania, e-mail:
aida$_{-}$toma@yahoo.com}
\address[label2]{``Gh. Mihoc - C. Iacob'' Institute of Mathematical
Statistics and Applied Mathematics, Calea 13 Septembrie 13,
Bucharest, Romania} \cortext[cor1]{Corresponding author.
Department of Applied Mathematics, Bucharest Academy of Economic
Studies, Pia\c ta Roman\u a 6, Bucharest, Romania. E-mail:
aida$_{-}$toma@yahoo.com, tel. 0040216340493}

\author[label3]{Samuela Leoni-Aubin}

\address[label3]{INSA Lyon, ICJ, 20, Rue Albert Einstein,
 69621 Villeurbanne Cedex, France,\\ e-mail: samuela.leoni@insa-lyon.fr }

\begin{abstract}The presence of outliers in financial asset
returns is a frequently occuring phenomenon and may lead to
unreliable mean-variance optimized portfolios. This fact is due to
the unbounded influence that outliers can have on the mean returns
and covariance estimators that are inputs in the optimization
procedure. In the present paper we consider new robust estimators
of location and covariance obtained by minimizing an empirical
version of a pseudodistance between the assumed model and the true
model underlying the data. We prove statistical properties of the
new mean and covariance matrix estimators, such as affine
equivariance, B-robustness and efficiency. These estimators can be
easily used in place of the classical estimators, thereby
providing robust optimized portfolios. A Monte Carlo simulation
study and an application to real data show the advantages of the
proposed approach.
\end{abstract}


\end{frontmatter}



\section{Introduction}Since Markowitz (1952) formulated the idea
of diversification of investments, the mean-variance approach has
been widely used in practice in asset allocation and portfolio
management, despite many sophisticated models proposed in
literature. On the other hand, some drawbacks of the standard
Markowitz approach are reported in literature (Michaud (1989)).
One of the critical weaknesses of the classical mean-variance
analysis is its lack of robustness. Since the classical estimators
of the mean and the covariance matrix, which are inputs in the
optimization procedure, are very sensitive to the presence of
gross errors or atypical events in data, the weights of the
resulted portfolio, which are outputs of this procedure, can be
drastically affected by these atypical data. This fact was proved
by Perret-Gentil and Victoria-Feser (2005) by using the influence
function approach. Also, some other recent papers underline this
idea and show that the large or small values of asset returns can
have an abnormally large influence on the estimations leading to
portfolio that are far to be optimal (Grossi and Laurini (2011)).
In order to remove this drawback and to construct portfolios not
overly affected by deviations of the data from the assumed model,
many methods have been proposed in literature. For an overview on
the methods used for robust portfolio optimization we refer to
Fabozzi et al. (2010). Among the methods which improve the
stability of portfolio weights by using robust estimators of the
mean and covariance, we recall those proposed by Vaz-de Melo and
Camara (2005) which use M-estimators, Perret-Gentil and
Victoria-Feser (2005) which use the translated biweight
S-estimator, Welsch and Zhou (2007) which use minimum covariance
determinant estimator and winsorization, DeMiguel and Nogales
(2009) which use both M- and S-estimators, Ferrari and Paterlini
(2010) which use Maximum L$_{q}$-Likelihood Estimators. These
contributions have the merit to consider the role of robust
estimation for improving the mean-variance portfolios. On the
other hand, it is known that traditional robust estimators suffer
dramatic looses in efficiency compared with the maximum likelihood
estimator. Therefore, a trade-off between robustness and
efficiency should be carefully analyzed.

Our contribution to robust portfolio optimization is developed
within a minimum pseudodistance framework. We can say that the
minimum pseudodistance methods for estimation take part to the
same cathegory with minimum divergence methods. The minimum
divergence estimators are defined by minimizing some appropriate
divergence between an assumed model and the true model underlying
the data. Depending on the choice of the divergence, minimum
divergence estimators can afford considerable robustness at
minimal loss of efficiency. However, the classical approaches
based on divergence minimization require nonparametric density
estimation, which can be problematic in multi-dimensional
settings. Some proposals to avoid the nonparametric density
estimation in minimum divergence estimation have been made by Basu
et al. (1998) and Broniatowski and Keziou (2009) and robustness
properties of their estimators have been studied by Toma and
Leoni-Aubin (2010), Toma and Broniatowski (2011).

In this paper we consider estimators of location and covariance
obtained by minimizing a family of pseudodistances. These
estimators have the advantages of not requiring any prior
smoothing and conciliate robustness with high efficiency, usually
requiring distinct techniques. The minimum pseudodistance
estimators have been introduced by Broniatowski et al. (2012) and
consist in minimization of an empirical version of a
pseudodistance between the assumed model and the true model
underlying the data. This method can be applied to any parametric
model, but in the present paper we focus on the multivariate
normal location-scale model. The behavior of the estimator depends
on a tuning positive parameter $\alpha$ which controls the
trade-off between robustness and efficiency. When the data are
consistent with normality and $\alpha\to 0$, the estimation method
corresponds to the maximum likelihood method (MLE) which is known
to have full asymptotic efficiency at the model. When $\alpha>0$,
the estimator gains robustness, while keeping high efficiency. The
new minimum pseudodistance estimators can be easily used in place
of the classical mean and covariance matrix estimators, thereby
providing robust and efficient mean-variance optimized portfolios.

The outline of the paper is as follows: In Section 2, we shortly
describe the Markowitz's mean-variance model whose inputs are
estimations of location and covariance of asset returns. The
minimum pseudodistance estimators of location and covariance are
introduced in Section 3. Here we prove theoretical properties of
these estimators, such as the affine equivariance and
B-robustness. We also determine the asymptotic covariance matrices
of the estimators and discuss the asymptotic relative efficiency.
The estimators of the portfolio weights together with their
properties are presented in Section 4. In Sections 5 and 6, a
Monte Carlo simulation study and then an application on real data
show the advantages of the new approach.
\section{Portfolio optimization model}
We consider a portfolio formed by $N$ financial assets. The
returns of the assets are characterized by the random vector
$X=(X_{1},\dots,X_{N})^{t}$, where $X_{i}$ denotes the random
variable associated to the return of the asset $i$, $i=1,\dots,N$.
Let $p=(p_{1},\dots,p_{N})^{t}$ be the vector of weights
associated to the portfolio, where $p_{i}$ represents the
proportion of the investor's capital invested in the asset $i$.
The total return of the portfolio is given by the random variable
\begin{equation*}
p^{t}X=p_{1}X_{1}+\dots+p_{N}X_{N}.
\end{equation*}

Supposing that the random vector $X$ follows a multivariate normal
distribution $\mathcal{N}_{N}(\mu,\Sigma)$, where $\mu$ is the
vector containing the mean returns of the assets and $\Sigma$ is
the covariance matrix of the returns of the assets, the mean of
the portfolio return can be written as $R(p)=p^{t}\mu$ and the
portfolio variance as $S(p)=p^{t}\Sigma p.$

The Markowitz approach for optimal portfolio selection consists in
solving the following optimization problem. For a given investor's
risk aversion $\lambda>0$, the mean-variance optimization selects
the portfolio $p^{*}$, solution of
\begin{equation*}
\arg\max_{p}\{R(p)-\frac{\lambda}{2}S(p)\}
\end{equation*}
with the constraint $p^{t}e_{N}=1$, $e_{N}$ being the $N\times1$
vector of ones. The set of optimal portfolios for all possible
values of the risk aversion parameter $\lambda$ defines the
mean-variance efficient frontier. The solution of the above
optimization problem is explicit and the optimal portfolio
weights, for a fixed value of $\lambda$, are given by
\begin{equation}\label{f1}
p^{*}=\frac{1}{\lambda}\Sigma^{-1}(\mu-\eta e_{N})
\end{equation}
where
\begin{equation*}
\eta=\frac{e_{N}^{t}\sum^{-1}\mu-\lambda}{e_{N}^{t}\sum^{-1}e_{N}}.
\end{equation*}
This is the case when short selling is allowed. When short selling
is not allowed, we have a supplementary constraint in the
optimization problem, namely all the weights $p_{i}$ are positive.

When the true parameters $\mu$ and $\Sigma$ and the portfolio
weights are all known, then we have the true efficient frontier.
An estimated efficient frontier can be obtained by using
estimators of the mean and covariance matrix. Throughout this
paper we denote by $\widehat{\mu}$ and $\widehat{\Sigma}$ the
estimators of the parameters $\mu$ and $\Sigma$, and by
$\widehat{p^{*}}$ the estimator of the optimal portfolio weights,
as resulting with (\ref{f1})
\begin{equation}\label{estwei}
\widehat{p^{*}}=\frac{1}{\lambda}\widehat{\Sigma}^{-1}\left[\widehat{\mu}-\frac{e_{N}^{t}\widehat{\Sigma}^{-1}\widehat{\mu}-\lambda}{e_{N}^{t}\widehat{\Sigma}^{-1}e_{N}}e_{N}\right].
\end{equation}
The mean and the covariance matrix of the returns are in practice
estimated by their sample counterparts, i.e. the maximum
likelihood estimators under the multivariate normal model. It is
known that, under normality, the maximum likelihood estimators are
the most efficient. However, in the presence of outlying
observations, the asymptotic bias of these estimators can be
arbitrarily large and this bias is induced to the corresponding
optimal portfolio weights. For this reason, $\mu$ and $\Sigma$
should be robustly estimated.

\section{Robust estimators of the location and covariance}
\subsection{Minimum pseudodistance estimators}

In the following, for the robust estimation of the parameters
$\mu$ and $\Sigma$ we consider minimum pseudodistance estimators.
For two probability measures $P$ and $Q$, admitting densities $p$,
respectively $q$ with respect to the Lebesgue measure $\lambda$,
the pseudodistances that we consider are defined through
\begin{equation*}
R_{\alpha}(P,Q):=\frac{1}{\alpha+1}\ln \int
p^{\alpha}\mathrm{d}P+\frac{1}{\alpha(\alpha+1)}\ln \int
q^{\alpha}\mathrm{d}Q-\frac{1}{\alpha}\ln \int
p^{\alpha}\mathrm{d}Q
\end{equation*}
for $\alpha>0$ and satisfy the limit relation
\begin{equation*}
R_{\alpha}(P,Q)\to R_{0}(P,Q):=\int \ln
\frac{q}{p}\mathrm{d}Q\;\;\text{for}\;\;\alpha\downarrow 0.
\end{equation*}
Note that $R_{0}(P,Q)$ coincides with the modified
Kullback-Leibler divergence.

Let $\mathcal{P}$ be a parametric model with parameter space
$\Theta\subset\mathbb{R}^{d}$ and assume that every probability
measure $P_{\theta}$ in $\mathcal{P}$ has a density $p_{\theta}$
with respect to the Lebesgue measure. The family of minimum
pseudodistance estimators of the unknown parameter $\theta_{0}$ is
obtained by replacing the hypothetical probability measure
$P_{\theta_{0}}$ in the pseudodistances
$R_{\alpha}(P_{\theta},P_{\theta_{0}})$ by the empirical measure
$P_{n}$ pertaining to the sample and then minimizing
$R_{\alpha}(P_{\theta},P_{n})$ with respect to $\theta$ on the
parameter space.

Let $X^{1},\dots,X^{T}$ be a sample on
$X\sim\mathcal{N}_{N}(\mu,\Sigma)$ and denote by
$\theta=(\mu,\Sigma)$ the parameter of interest. A minimum
pseudodistance estimator
$\widehat{\theta}=(\widehat{\mu},\widehat{\Sigma})$ of $\theta$ is
defined by
\begin{equation*}
\widehat{\theta}:=\arg\inf_{\theta}R_{\alpha}(P_{\theta},P_{n})
\end{equation*}
which can be written equivalently as
\begin{equation}\label{mpe}
\widehat{\theta}=\left\{
\begin{array}
[c]{ll}%
\arg\sup_{\theta}\frac{1}{TC_{\alpha}(\theta)}\sum_{i=1}^{T}p_{\theta}^{\alpha}(X^{i})
& \text{ \ \ \ }\mbox{if}\ \alpha>0\medskip\\
\arg\sup_{\theta}\frac{{ 1}}{{
T}}\sum_{i=1}^{T}\ln p_{{\theta}}(X^{i}) & \text{ \ \ \ }%
\mbox{if}\text{ }\alpha=0
\end{array}
\right.
\end{equation}
where $p_{\theta}$ is the $N$-variate normal density
\begin{equation*}
p_{\theta}(x)=\left(\frac{1}{2\pi}\right)^{N/2}\sqrt{\det
\Sigma^{-1}}\exp\left(-\frac{1}{2}(x-\mu)^{t}\Sigma^{-1}(x-\mu)\right)
\end{equation*}
and $C_{\alpha}(\theta)=\left(\int
p_{\theta}^{\alpha+1}d\lambda\right)^{\frac{\alpha}{\alpha+1}}$.
Note that, the choice $\alpha=0$ leads to the definition of the
classical MLE. Throughout the paper we will also use the notation
$\|x-\mu\|_{\Sigma^{-1}}:=[(x-\mu)^{t}\Sigma^{-1}(x-\mu)]^{1/2}$.
A simple calculation shows that
\begin{equation*}
C_{\alpha}(\theta)=\frac{\left(\frac{1}{2\pi}\right)^{\frac{N\alpha^2}{2(\alpha+1)}}(\sqrt{\det
\Sigma^{-1}
})^{\frac{\alpha^{2}}{\alpha+1}}}{(\sqrt{\alpha+1})^{\frac{N
\alpha}{\alpha+1}}}
\end{equation*}
and then, for $\alpha>0$ the minimum pseudodistance estimator
(\ref{mpe}) can be expressed as
\begin{equation*}
\widehat{\theta}=\arg\sup_{\theta}(\sqrt{\det
\Sigma^{-1}})^{\frac{\alpha}{\alpha+1}}\sum_{i=1}^{T}\exp\left(-\frac{\alpha}{2}\|X^{i}-\mu\|_{\Sigma^{-1}}^{2}\right).
\end{equation*}

By direct differentiation with respect to $\mu$ and $\Sigma$, we
see that the estimators of these parameters are solutions of the
system
\begin{eqnarray}
\mu&=&\sum_{i=1}^{T}\frac{\exp(-\frac{\alpha}{2}\|X^{i}-\mu\|_{\Sigma^{-1}}^{2})}{\sum_{i=1}^{T}\exp(-\frac{\alpha}{2}\|X^{i}-\mu\|_{\Sigma^{-1}}^{2})}X^{i}\label{sys1}\\
\Sigma&=&\sum_{i=1}^{T}\frac{(\alpha+1)\exp\left(-\frac{\alpha}{2}\|X^{i}-\mu\|_{\Sigma^{-1}}^{2}\right)}{\sum_{i=1}^{T}\exp(-\frac{\alpha}{2}\|X^{i}-\mu\|_{\Sigma^{-1}}^{2})}(X^{i}-\mu)(X^{i}-\mu)^{t}\label{sys2}.
\end{eqnarray}

In order to compute $\widehat{\mu}$ and $\widehat{\Sigma}$ we use
a reweighting algorithm which we describe in Section \ref{simul}.
\subsection{Affine equivariance}
The location and dispersion estimators defined above are affine
equivariant. More precisely, if $\widehat{\mu}(\mathbf{X})$ and
$\widehat{\Sigma}(\mathbf{X})$ are estimators corresponding to a
sample $\mathbf{X}=(X^{1},\dots,X^{T})$, then
\begin{eqnarray}
\widehat{\mu}(A\mathbf{X}+b)&=&A\widehat{\mu}(\mathbf{X})+b\label{AE1}\\
\widehat{\Sigma}(A\mathbf{X}+b)&=&A\widehat{\Sigma}(\mathbf{X})A^{t}\label{AE2}
\end{eqnarray}
for any $N\times N$ nonsingular matrix $A$ and any $b\in
\mathbb{R}^{N}$.

Indeed, let $A$ be a $N\times N$ nonsingular matrix, $b\in
\mathbb{R}^{N}$ and $\mathbf{Y}=(Y^{1},\dots,Y^{T})$ with
$Y^{i}:=AX^{i}+b$. The estimators $\widehat{\mu}(\mathbf{Y})$ and
$\widehat{\Sigma}(\mathbf{Y})$ are solutions of the system
obtained from (\ref{sys1}) and (\ref{sys2}) by replacing $X^{i}$
with $Y^{i}$. Then, by replacing  $Y^{i}$ with $AX^{i}+b$, we find
$\widehat{\mu}(\mathbf{X})=A^{-1}(\widehat{\mu}(\mathbf{Y})-b)$
and
$\widehat{\Sigma}(\mathbf{X})=A^{-1}\widehat{\Sigma}(\mathbf{Y})(A^{t})^{-1}$.
Hence, (\ref{AE1}) and (\ref{AE2}) hold.

\subsection{Influence functions}

A fundamental tool used for studying statistical robustness is the
influence function. The influence function is useful to determine
analytically and numerically the stability properties of an
estimator in case of model misspecification. Recall that, a map
$T$ defined on a set of probability measures and parameter space
valued is a statistical functional corresponding to an estimator
$\widehat{\theta}_{n}$ of the parameter $\theta$, if
$\widehat{\theta}_{n}=T(P_{n})$, where $P_{n}$ is the empirical
measure associated to the sample. As it is known, the influence
function of $T$ at $P_{\theta}$ is defined by
\begin{equation*}
\mathrm{IF}(x;T,P_{\theta}):=\left.\frac{\partial
T(\widetilde{P}_{\varepsilon
x})}{\partial\varepsilon}\right|_{\varepsilon=0}
\end{equation*}
where $\widetilde{P}_{\varepsilon
x}:=(1-\varepsilon)P_{\theta}+\varepsilon \delta_{x},$
$\varepsilon>0$, $\delta_{x}$ being the Dirac measure putting all
mass at $x$. Whenever the influence function is bounded with
respect to $x$ the corresponding estimator is called robust.

The statistical functionals associated to the minimum
pseudodistance estimators of $\mu$ and $\Sigma$ are $\mu(P)$ and
$\Sigma(P)$ defined by the solutions of the system
\begin{eqnarray*}
&&\int
(x-\mu)\exp\left(-\frac{\alpha}{2}\|x-\mu\|^{2}_{\Sigma^{-1}}\right)dP(x)=0\\
&&\int
\left[(x-\mu)(x-\mu)^{t}\exp\left(-\frac{\alpha}{2}\|x-\mu\|^{2}_{\Sigma^{-1}}\right)-\frac{1}{\alpha+1}\Sigma\exp\left(-\frac{\alpha}{2}\|x-\mu\|^{2}_{\Sigma^{-1}}\right)\right]dP(x)=0.
\end{eqnarray*}
This system can be rewritten under the form
\begin{eqnarray}
&&\int w_{1}(\|x-\mu\|_{\Sigma^{-1}})(x-\mu)dP(x)=0\label{f44}\\
&&\int
\left[\frac{w_{2}(\|x-\mu\|_{\Sigma^{-1}})}{\|x-\mu\|_{\Sigma^{-1}}}(x-\mu)(x-\mu)^{t}-w_{3}(\|x-\mu\|_{\Sigma^{-1}})\Sigma\right]
dP(x)=0\label{f4}
\end{eqnarray}
where
\begin{equation}\label{f5}
w_{1}(t)=\exp \left(-\frac{\alpha}{2}t^{2}\right)\;, w_{2}(t)=\exp
\left(-\frac{\alpha}{2}t^{2}\right)t^{2},\;
w_{3}(t)=\frac{1}{\alpha+1}\exp
\left(-\frac{\alpha}{2}t^{2}\right).
\end{equation}

\begin{figure}[t]
\begin{tabular}{ c c}
 \hspace{-10mm}\includegraphics[width=8cm,height=8cm]{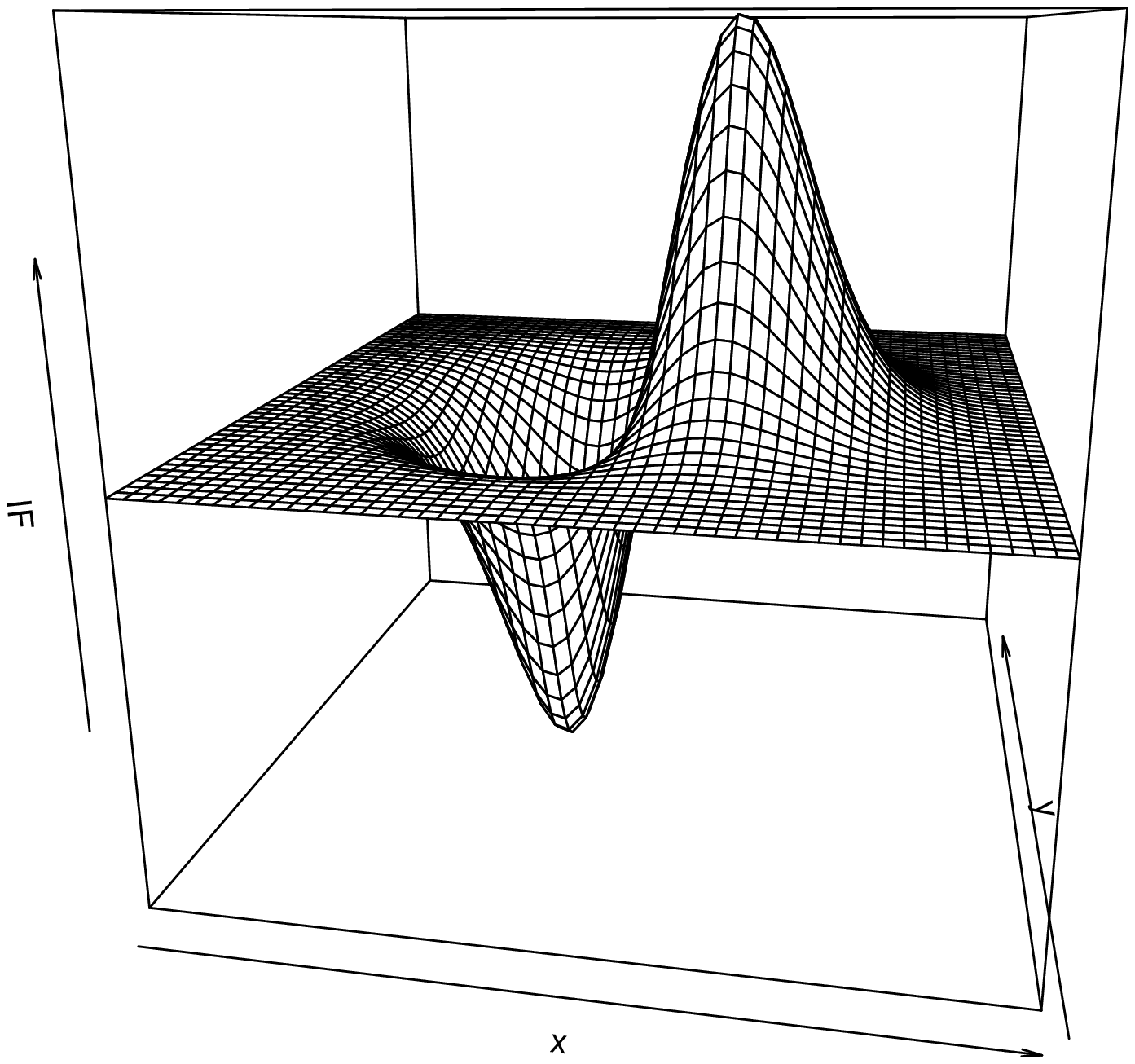}&
 \includegraphics[width=8cm,height=8cm]{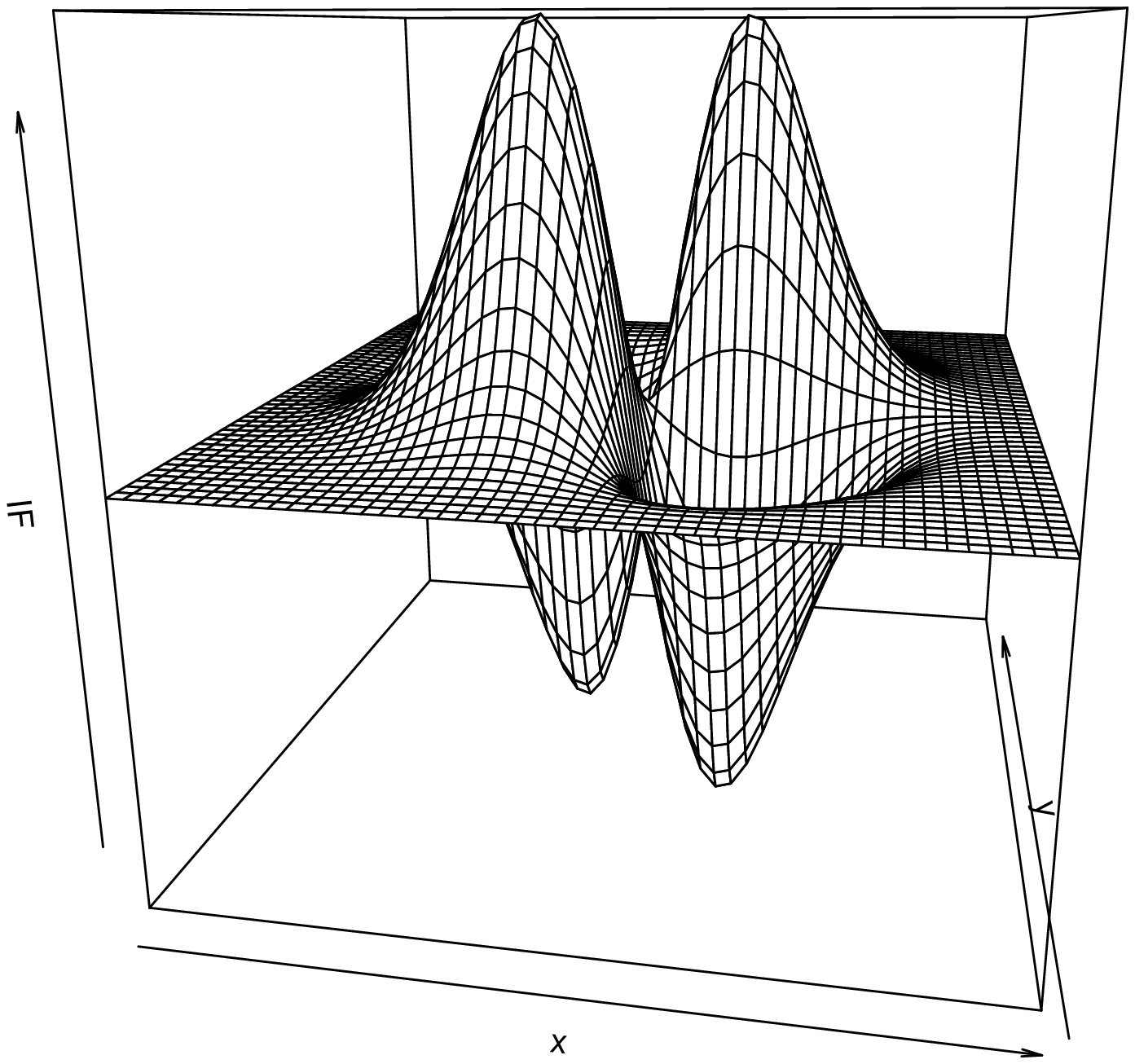}
\end{tabular}
\vspace{-2mm}\caption{The influence function for the first
component of the minimum pseudodistance estimator of the mean
(left hand side) and the influence function for the ex diagonal
component of the minimum pseudodistance estimator of the
covariance matrix (right hand side). $P_{0}$ is the bivariate
standard normal law and $\alpha=0.5$}\label{fig1}
\end{figure}

We note that the solutions of the system given by (\ref{f44}) and
(\ref{f4}), when $w_{1},w_{2},w_{3}$ are arbitrary weight
functions, define statistical functionals of general M-estimators
of $(\mu, \Sigma)$ (see Huber (1977), Jaupi and Saporta (1993)).
According to the results presented by Jaupi and Saporta (1993),
the influence functions for general M-estimators of $\mu$ and
$\Sigma$ are given by
\begin{eqnarray}
\mathrm{IF}(x;\mu,P_{\mu,\Sigma})&=&(x-\mu)w_{\mu}(\|x-\mu\|_{\Sigma^{-1}})\label{f6}\\
\mathrm{IF}(x;\Sigma,P_{\mu,\Sigma})&=&(x-\mu)(x-\mu)^{t}w_{\eta}(\|x-\mu\|_{\Sigma^{-1}})-\Sigma
w_{\delta}(\|x-\mu\|_{\Sigma^{-1}})\label{f7}
\end{eqnarray}
where
\begin{eqnarray*}
w_{\mu}(\|x-\mu\|_{\Sigma^{-1}})&=&\frac{w_{1}(\|x-\mu\|_{\Sigma^{-1}})}{E_{P_{0}}\left[w_{1}(\|y\|)+\frac{1}{N}w_{1}^{\prime}(\|y\|)\|y\|\right]}\\
w_{\eta}(\|x-\mu\|_{\Sigma^{-1}})&=&\frac{N(N+2)w_{2}(\|x-\mu\|_{\Sigma^{-1}})}{\|x-\mu\|^{2}_{\Sigma^{-1}}E_{P_{0}}\left[Nw_{2}(\|y\|)+w_{2}^{\prime}(\|y\|)\|y\|\right]}\\
w_{\delta}(\|x-\mu\|_{\Sigma^{-1}})&=&\frac{Nw_{3}(\|x-\mu\|_{\Sigma^{-1}})-2w_{2}(\|x-\mu\|_{\Sigma^{-1}})}{E_{P_{0}}\left[w_{2}^{\prime}(\|y\|)\|y\|-Nw_{3}^{\prime}(\|y\|)\|y\|\right]}+\\
&&+\frac{(N+2)w_{2}(\|x-\mu\|_{\Sigma^{-1}})}{E_{P_{0}}\left[Nw_{2}(\|y\|)+w_{2}^{\prime}(\|y\|)\|y\|\right]}
\end{eqnarray*}
$P_{0}$ denoting the probability measure associate to the
$N$-variate standard normal distribution and $\|\cdot\|$ the
Euclidian norm.

For the weight functions $w_{1},w_{2},w_{3}$ from (\ref{f5}),
corresponding to the minimum pseudodistance estimators, we get
\begin{eqnarray}
w_{\mu}(t)&=&(\sqrt{\alpha+1})^{N+2}\exp\left(-\frac{\alpha}{2}t^{2}\right)\label{wmu}\\
w_{\eta}(t)&=&(\sqrt{\alpha+1})^{N+4}\exp\left(-\frac{\alpha}{2}t^{2}\right)\label{weta}\\
w_{\delta}(t)&=&(\sqrt{\alpha+1})^{N+2}\exp\left(-\frac{\alpha}{2}t^{2}\right)\label{wdelta}
\end{eqnarray}
and replacing in (\ref{f6}) and in (\ref{f7}) we obtain
\begin{eqnarray}
\mathrm{IF}(x;\mu,P_{\mu,\Sigma})&=&(\sqrt{\alpha+1})^{N+2}(x-\mu)\exp\left(-\frac{\alpha}{2}\|x-\mu\|^{2}_{\Sigma^{-1}}\right)\label{IF11}\\
\mathrm{IF}(x;\Sigma,P_{\mu,\Sigma})&=&(\sqrt{\alpha+1})^{N+4}\left[(x-\mu)(x-\mu)^{t}-\frac{1}{\alpha+1}\Sigma\right]\exp\left(-\frac{\alpha}{2}\|x-\mu\|^{2}_{\Sigma^{-1}}\right)\label{IF22}.
\end{eqnarray}

Both influence functions are bounded with respect to $x$.
Therefore the minimum pseudodistance estimators of $\mu$ and
$\Sigma$ are robust. In Figure \ref{fig1} we represent the
influence function for the first component of the minimum
pseudodistance estimator of the mean, respectively the influence
function for the ex diagonal component of the minimum
pseudodistance estimator of the covariance matrix. For both
representations, we considered $P_{0}$ the bivariate standard
normal low and we chosed $\alpha=0.5$.

\subsection{Asymptotic normality}
For general parametric models, the minimum pseudodistance
estimators are asymptotically normal distributed (see Broniatowski
et al. (2012)). In this section, we derive the asymptotic
covariance matrices of the mean and the covariance matrix minimum
pseudodistance estimators. We adopt the influence function
approach and make use of the general results for affine
equivariant location and dispersion M-estimators as presented in
Gervini (2002) and Hampel et al. (1986).

When the observations correspond to the standard $N$-variate
normal law $P_{0}$, under appropriate conditions, $\widehat{\mu}$
is asymptotically normal distributed with the asymptotic
covariance matrix
\begin{equation}
V(\mu,P_{0})=\mathrm{E}_{P_{0}}\{\mathrm{IF}(Z;\mu,P_{0})\mathrm{IF}(Z;\mu,P_{0})^{t}\}=d_{\mu}I\label{IFmu0}
\end{equation}
where
$d_{\mu}:=\mathrm{E}_{P_{0}}\{\|Z\|^{2}w_{\mu}^{2}(\|Z\|)\}/N$ and
$I$ is the identity matrix. Formula (\ref{IFmu0}) has been
established by Gervini (2002) for general affine equivariant
location M-estimators. The estimator $\widehat{\mu}$ belongs to
this class. For the weight $w_{\mu}$ from (\ref{wmu}) we get
$d_{\mu}=(\alpha+1)^{N+2}/(\sqrt{2\alpha+1})^{N+2}$, hence the
asymptotic covariance matrix of the minimum pseudodistance
estimator $\widehat{\mu}$ is
\begin{equation*}
V(\mu,P_{0})=\frac{(\alpha+1)^{N+2}}{(\sqrt{2\alpha+1})^{N+2}}I.
\end{equation*}
When the observations correspond to the normal law
$P_{\mu,\Sigma}$, the asymptotic covariance matrix of
$\widehat{\mu}$ is given by
\begin{equation}
V(\mu,P_{\mu,\Sigma})=E_{P_{\mu,\Sigma}}\{\mathrm{IF}(X;\mu,P_{\mu,\Sigma})\mathrm{IF}(X;\mu,P_{\mu,\Sigma})^{t}\}=d_{\mu}\Sigma=\frac{(\alpha+1)^{N+2}}{(\sqrt{2\alpha+1})^{N+2}}\Sigma.\label{ifmuP0}
\end{equation}

Similar results hold for $\mathrm{vecs}(\widehat{\Sigma})$, where
$\mathrm{vecs}$ is the operation that stacks the $N+N(N-1)/2$
non-redundant elements of $\Sigma$ into a vector, as follows:
$\mathrm{vecs}(\Sigma):=(\sigma_{11}/\sqrt{2},\dots,\sigma_{NN}/\sqrt{2},\sigma_{21},\sigma_{31},\dots,\sigma_{N,N-1})^{t}$.
According to the results of Gervini (2002), when the observations
come from the $N$-variate standard normal law $P_{0}$, the
asymptotic covariance matrix corresponding to an affine
equivariant M-estimator of the covariance matrix is given by
\begin{eqnarray*}
V(\Sigma,P_{0})&=&\mathrm{E}_{P_{0}}\{\mathrm{vecs}\mathrm{IF}(Z;\Sigma,P_{0})\mathrm{vecs}\mathrm{IF}(Z;\Sigma,P_{0})^{t}\}\\
&=&d_{\eta}(I-\frac{1}{N}ww^{t})+d_{\tau}\cdot\frac{1}{N}ww^{t}
\end{eqnarray*}
where $w^{t}:=(e^{t}_{N},\mathbf{0}_{N(N-1)/2}^{t})$,
$d_{\eta}:=\mathrm{E}_{P_{0}}\{\|Z\|^{4}w_{\eta}^{2}(\|Z\|)\}/(N(N+2))$
and $d_{\tau}:=\mathrm{E}_{P_{0}}\{w_{\tau}^{2}(\|Z\|)\}/(2N)$
with $w_{\tau}:=t^{2}w_{\eta}(t)-Nw_{\delta}(t)$, $w_{\eta}$,
$w_{\delta}$ and $w_{\tau}$ being specific to the M-estimator in
question.

In our case, $w_{\eta}$ and $w_{\delta}$ are given by (\ref{weta})
and (\ref{wdelta}), hence
\begin{equation*}
w_{\tau}(t)=(\sqrt{\alpha+1})^{N+4}\left[t^2-\frac{N}{\alpha+1}\right]\exp\left(-\frac{\alpha}{2}t^2\right).
\end{equation*}
After some calculation, we obtain
\begin{equation*}
d_{\eta}=\left(\frac{\alpha+1}{\sqrt{2\alpha+1}}\right)^{N+4}\;\text{and}\;
d_{\tau}=\frac{N\alpha^2(\alpha+1)^{N+2}}{2(\sqrt{2\alpha+1})^{N+4}}+\left(\frac{\alpha+1}{\sqrt{2\alpha+1}}\right)^{N+4},
\end{equation*}
therefore,
\begin{equation}\label{VP0}
V(\Sigma,P_{0})=\left(\frac{\alpha+1}{\sqrt{2\alpha+1}}\right)^{N+4}I+\frac{\alpha^2(\alpha+1)^{N+2}}{2(\sqrt{2\alpha+1})^{N+4}}ww^{t}.
\end{equation}

When the observations correspond to the law $P_{\mu,\Sigma}$, the
asymptotic covariance matrix of $\mathrm{vecs}(\widehat{\Sigma})$
can be established by using the formula from Hampel et al. (1986)
p.282, which in our notations writes as follows
\begin{equation}\label{VPmS}
V(\Sigma,P_{\mu,\Sigma})=\left[\frac{\partial[\Sigma^{\frac{1}{2}}S\Sigma^{\frac{1}{2}}]}{\partial
\mathrm{vecs}S
}\right]V(\Sigma,P_{0})\left[\frac{\partial[\Sigma^{\frac{1}{2}}S\Sigma^{\frac{1}{2}}]}{\partial
\mathrm{vecs}\Sigma }\right]^{t}.
\end{equation}
According to Hampel et al. (1986) p.272, for a given $N\times N$
matrix $\Sigma_{*}$, it holds
\begin{equation*}
\left[\frac{\partial[\Sigma^{\frac{1}{2}}S\Sigma^{\frac{1}{2}}]}{\partial
\mathrm{vecs}S
}\right]\mathrm{vecs}\Sigma_{*}=\mathrm{vecs}(\Sigma^{\frac{1}{2}}\Sigma_{*}\Sigma^{\frac{1}{2}}).
\end{equation*}
Particularly,
\begin{equation}\label{pass}
\left[\frac{\partial[\Sigma^{\frac{1}{2}}S\Sigma^{\frac{1}{2}}]}{\partial
\mathrm{vecs}S }\right]\mathrm{vecs}I=\mathrm{vecs}\Sigma.
\end{equation}
Note that $w=\sqrt{2}\mathrm{vecs}I$ and combining (\ref{VPmS}),
(\ref{VP0}) and (\ref{pass}), we get
\begin{equation*}
V(\Sigma,P_{\mu,\Sigma})=\left(\frac{\alpha+1}{\sqrt{2\alpha+1}}\right)^{N+4}\left[\frac{\partial[\Sigma^{\frac{1}{2}}S\Sigma^{\frac{1}{2}}]}{\partial
\mathrm{vecs}S
}\right]\left[\frac{\partial[\Sigma^{\frac{1}{2}}S\Sigma^{\frac{1}{2}}]}{\partial
\mathrm{vecs}S
}\right]^{t}+\frac{\alpha^2(\alpha+1)^{N+2}}{(\sqrt{2\alpha+1})^{N+4}}\mathrm{vecs}\Sigma(\mathrm{vecs}\Sigma)^{t}.
\end{equation*}
For symmetry reasons, the minimum pseudodistance location and
covariance estimators are asymptotically uncorrelated and hence
asymptotically independent. This is valid for location and
covariance M-estimators in general, as it is underlined in various
articles, for example in Huber (1977).

\subsection{Asymptotic relative efficiency}
In order to assess the efficiency of the proposed estimators with
respect to that of the MLE, we adopt as measure the asymptotic
relative efficiency (ARE). For a parameter $\theta$ taking values
in $\mathbb{R}^{d}$ and an estimator $\widehat{\theta}$ which is
asymptotically $d$-variate normal with mean $\theta$ and
nonsingular covariance matrix $V(\theta,P)$, the asymptotic
relative efficiency with respect to that of the MLE is defined as
\begin{equation*}
\mathrm{ARE}(\widehat{\theta},P)=\left(\frac{\det
V_{0}(\theta,P)}{\det V(\theta,P) }\right)^{1/d},
\end{equation*}
$V_{0}(\theta,P)$ being the asymptotic covariance matrix of the
MLE of $\theta$ when the observations follow the law $P$ (see
Serfling (2011)). Although the asymptotically most efficient
estimator is given by the MLE, the particular MLE can be
drastically inefficient when the underlying distribution departs
even a little bit from the assumed nominal distribution. Therefore
the trade-off between robustness and efficiency should be
carefully analyzed.

Due to the asymptotic independence of the mean and the covariance
matrix minimum pseudodistance estimators, the asymptotic relative
efficiency of
$\widehat{\theta}=(\widehat{\mu}^{t},\mathrm{vecs}(\widehat{\Sigma})^{t})^{t}$
can be expressed as
\begin{equation}
\mathrm{ARE}(\widehat{\theta},P_{\mu,\Sigma})=\left(\frac{\det
V_{0}(\theta,P_{\mu,\Sigma})}{\det
V(\theta,P_{\mu,\Sigma})}\right)^{\frac{2}{N(N+3)}}=\left(\frac{\det
V_{0}(\mu,P_{\mu,\Sigma})\det V_{0}(\Sigma,P_{\mu,\Sigma})}{\det
V(\mu,P_{\mu,\Sigma})\det
V(\Sigma,P_{\mu,\Sigma})}\right)^{\frac{2}{N(N+3)}}.\label{are0}
\end{equation}
Using (\ref{ifmuP0}) and (\ref{VPmS}), formula (\ref{are0}) can be
written as
\begin{equation*}
\mathrm{ARE}(\widehat{\theta},P_{\mu,\Sigma})=\left(\frac{\det
V_{0}(\mu,P_{0})\det V_{0}(\Sigma,P_{0})}{\det V(\mu,P_{0})\det
V(\Sigma,P_{0})}\right)^{\frac{2}{N(N+3)}}.
\end{equation*}
A direct calculation shows that
\begin{eqnarray*}
\det
V(\mu,P_{0})&=&\left(\frac{\alpha+1}{\sqrt{2\alpha+1}}\right)^{N(N+2)}\\
\det
V(\Sigma,P_{0})&=&\left(\frac{\alpha+1}{\sqrt{2\alpha+1}}\right)^{\frac{N(N+1)(N+4)}{2}}\left(1+\frac{N\alpha^2}{2(\alpha+1)^2}\right).
\end{eqnarray*}
Particularly, for $\alpha=0$, we find the similar quantities for
the MLE, namely $\det V_{0}(\mu,P_{0})=1$ and $\det
V_{0}(\Sigma,P_{0})=1$. Hence
\begin{equation}
\mathrm{ARE}(\widehat{\theta},P_{\mu,\Sigma})=\frac{1}{\left(\frac{\alpha+1}{\sqrt{2\alpha+1}}\right)^{\frac{N^{2}+7N+8}{N+3}}\left(1+\frac{N\alpha^2}{2(\alpha+1)^2}\right)^{\frac{2}{N(N+3)}}}.\label{AREth}
\end{equation}
Note that, for fixed $N$ and $\alpha$,
$\mathrm{ARE}(\widehat{\theta},P_{\mu,\Sigma})$ is the same,
whatever $\mu$ or $\Sigma$.

\begin{table}[ptb]
\begin{center}
\begin{tabular}{ccccccc}
  \hline \vspace{1mm} N & $\alpha=0$ & $\alpha=0.1$ & $\alpha=0.2$ & $\alpha=0.5$ & $\alpha=0.75$ & $\alpha=1$
   \\\hline
  1 & 1 & 0.98151 & 0.93871  & 0.76904 & 0.63774 & 0.53033  \\
  2 & 1 & 0.97704 & 0.92429  & 0.72086 & 0.57042 & 0.45266  \\
  3 & 1 & 0.97273 & 0.91051  & 0.67698 & 0.51187 & 0.38814  \\
  4 & 1 & 0.96851 & 0.89718  & 0.63647 & 0.46018 & 0.33370  \\
  5 & 1 & 0.96435 & 0.88419  & 0.59879 & 0.41420 & 0.28738  \\
  6 & 1 & 0.96025 & 0.87148  & 0.56360 & 0.37311 & 0.24778  \\
  7 & 1 & 0.95619 & 0.85902  & 0.53065 & 0.33629 & 0.21380  \\
  8 & 1 & 0.95215 & 0.84679  & 0.49975 & 0.30322 & 0.18460  \\
  9 & 1 & 0.94815 & 0.83477  & 0.47073 & 0.27350 & 0.15946  \\
  10 & 1 & 0.94418 & 0.82294 & 0.44345 & 0.24674 & 0.13779  \\ \hline
\end{tabular}
\caption{Asymptotic relative efficiency of the minimum
pseudodistance estimators}\label{table1}
\end{center}
\end{table}

In Table \ref{table1} values of the asymptotic relative efficiency
(\ref{AREth}) are given. As it can be seen, when $N$ or $\alpha$
increases, the asymptotic relative efficiency
$\mathrm{ARE}(\widehat{\theta},P_{\mu,\Sigma})$ decreases.
Therefore, values of $\alpha$ close to zero will provide high
efficiency and in the meantime the robustness of the estimation
procedure.

\section{The estimator of the optimal portfolio weights}

We consider the estimator $\widehat{p^{*}}$ of the optimal
portfolio weights, as given by (\ref{estwei}), with
$\widehat{\mu}$ and $\widehat{\Sigma}$ minimum pseudodistance
estimators.

The influence function of the estimator $\widehat{p^{*}}$ is
proportional to the influence functions of the estimators
$\widehat{\mu}$ and $\widehat{\Sigma}$. More precisely,
\begin{eqnarray}
\mathrm{IF}(x;p^{*},P_{\mu,\Sigma})&=&
-\Sigma^{-1}\mathrm{IF}(x;\Sigma,P_{\mu,\Sigma})p^{*}+\frac{1}{\lambda}\Sigma^{-1}\left\{\mathrm{IF}(x;\mu,P_{\mu,\Sigma})+\right.\nonumber\\
&&+\frac{e_{N}^{t}\Sigma^{-1}[\mathrm{IF}(x;\Sigma,P_{\mu,\Sigma})\Sigma^{-1}\mu-\mathrm{IF}(x;\mu,P_{\mu,\Sigma})]e_{N}}{e_{N}^{t}\Sigma^{-1}e_{N}}+\nonumber\\
&&\left.+\frac{(e_{N}^{t}\Sigma^{-1}\mathrm{IF}(x;\Sigma,P_{\mu,\Sigma})\Sigma^{-1}e_{N})(e_{N}^{t}\Sigma^{-1}\mu-\lambda)e_{N}}{(e_{N}^{t}\Sigma^{-1}e_{N})^{2}}\right\}
\label{f3}\end{eqnarray}
 where $\mathrm{IF}(x;\mu,P_{\mu,\Sigma})$
and $\mathrm{IF}(x;\Sigma,P_{\mu,\Sigma})$ are those from
(\ref{IF11}) and (\ref{IF22}). This formula is obtained by
considering the statistical functional associated to the optimal
portfolio weights,
\begin{equation*}
p^{*}(P)=\frac{1}{\lambda}\Sigma^{-1}(P)\left[\mu(P)-\frac{e_{N}^{t}\Sigma^{-1}(P)\mu(P)-\lambda}{e_{N}^{t}\Sigma^{-1}(P)e_{N}}e_{N}\right]
\end{equation*}
where $\Sigma^{-1}(P)$ denotes the statistical functional
corresponding to $\widehat{\Sigma}^{-1}$, and then deriving the
influence function, taking also into account that
\begin{equation*}
\mathrm{IF}(x;
\Sigma^{-1},P_{\mu,\Sigma})=-\Sigma^{-1}\mathrm{IF}(x;
\Sigma,P_{\mu,\Sigma})\Sigma^{-1}.
\end{equation*}
On the basis of the direct proportionality between the influence
function $\mathrm{IF}(x;p^{*},P_{\mu,\Sigma})$ and the influence
functions $\mathrm{IF}(x;\mu,P_{\mu,\Sigma})$ and
$\mathrm{IF}(x;\Sigma,P_{\mu,\Sigma})$, we deduce that the global
robustness of $\widehat{\mu}$ and $\widehat{\Sigma}$ is
transferred to the plug-in estimator $\widehat{p^{*}}$.

On the other hand, by using the multivariate Delta method, the
asymptotic normality of $\widehat{p^{*}}$ is kept, as well. Given
the i.i.d. observations $X^{1},\dots,X^{T}$ from $P_{\mu,\Sigma}$,
since $\widehat{\mu}$ and $\mathrm{vecs}(\widehat{\Sigma})$ are
asymptotically normal and the function
\begin{equation*}
h(\theta)=\frac{1}{\lambda}\Sigma^{-1}(\mu-\eta e_{N})
\end{equation*}
with $\theta=(\mu^{t},(\mathrm{vecs}\Sigma)^{t})^{t}$ is
differentiable, by applying the multivariate Delta method, it
holds
\begin{equation*}
\sqrt{n}(\widehat{p^{*}}-p^{*})\to
\mathcal{N}(0,V(p^{*},P_{\mu,\Sigma}))
\end{equation*}
where
$V(p^{*},P_{\mu,\Sigma})=\mathrm{D}h(\theta)V(\theta,P_{\mu,\Sigma})\mathrm{D}h(\theta)^{t}$,
$\mathrm{D}h(\theta)$ being the differential of $h$ in $\theta$
and
\begin{equation*}
V(\theta,P_{\mu,\Sigma})=\begin{pmatrix}
  V(\mu,P_{\mu,\Sigma}) & 0 \\
  0 & V(\Sigma,P_{\mu,\Sigma})
\end{pmatrix}.
\end{equation*}
\section{Monte Carlo simulations}\label{simul}We performed Monte
Carlo simulations in order to assess the performance of the
minimum pseudodistance estimators of the mean and covariance
matrix, for both contaminated and non-contaminated data. In this
study, we considered the multivariate normal distribution
$\mathcal{N}_{N}(\mu_{0},\Sigma_{0})$, with $\mu_{0}=\mathbf{0}$
and $\Sigma_{0}$ a $N\times N$ matrix with variances equal to 1
and covariances all equal to 0.2. We generated samples of size $T$
in which about $(1-\varepsilon)T$ observations are from $
\mathcal{N}_{N}(\mu_{0},\Sigma_{0})$, while a smaller portion
$\varepsilon T$ is from the contaminating distribution
$\mathcal{N}_{N}(\mu_{c},\Sigma_{c})$ with $\mu_{c}=\mathbf{-4}$
and $\Sigma_{c}=4\Sigma_{0}$. We considered $N\in\{2,5,10,20\}$
and $\varepsilon\in\{0,0.05,0.1,0.2\}$. For each setting, we
generated 1000 samples and for each sample we computed minimum
pseudodistance estimates $\widehat{\mu}$ and $\widehat{\Sigma}$
corresponding to $\alpha\in\{0,0.1,0.2,0.5,0.75,1\}$.

The estimates $\widehat{\mu}$ and $\widehat{\Sigma}$, which are
solutions of the system of equations (\ref{sys1}) and
(\ref{sys2}), were obtained using the following reweighting
algorithm.

Let $s\in \{0,1,\dots,s^{*}\}$ denotes the iteration step.

1. If $s=0$

$\mu^{(s)}$ and $\Sigma^{(s)}$ are set to be initial estimates of
location and scale;

2. For $0<s<s^{*}$,
\begin{eqnarray*}
\mu^{(s)}&=&\sum_{i=1}^{T}w_{i}^{(s-1)}X^{i}\\
\Sigma^{(s)}&=&\sum_{i=1}^{T}(\alpha+1)w_{i}^{(s-1)}(X^{i}-\mu^{(s)})(X^{i}-\mu^{(s)})^{t}
\end{eqnarray*}
where
\begin{equation*}
w_{i}^{(s)}=\frac{\exp\left(-\frac{\alpha}{2}(X^{i}-\mu^{(s)})^{t}(\Sigma^{(s)})^{-1}(X^{i}-\mu^{(s)})\right)}{\sum_{i=1}^{T}\exp\left(-\frac{\alpha}{2}(X^{i}-\mu^{(s)})^{t}(\Sigma^{(s)})^{-1}(X^{i}-\mu^{(s)})\right)}.
\end{equation*}
At step 1, we used maximum likelihood estimates as initial
estimates of location and covariance. For details on general
convergence behavior of reweighting algorithms we refer to Arslan
(2004). If $\alpha>0$, the above procedure associates low weights
to the observations that disagree sensibly with the model. If
$\alpha=0$, all the observations receive the same weight and the
estimators are the maximum likelihood ones, defined through
\begin{eqnarray*}
\widehat{\mu}_{ML}&=&\frac{1}{T}\sum_{i=1}^{T}X^{i}\\
\widehat{\Sigma}_{ML}&=&\frac{1}{T}\sum_{i=1}^{T}(X^{i}-\widehat{\mu}_{ML})(X^{i}-\widehat{\mu}_{ML})^{t}.
\end{eqnarray*}
\begin{center}
\begin{table}[ptb]
\begin{tabular}{cccccccc}
  \hline
  &&&&$\varepsilon=0\%$&&&\\
  \hline \vspace{1mm}
  N&& $\alpha=0$ & $\alpha=0.1$ & $\alpha=0.2$ & $\alpha=0.5$ & $\alpha=0.75$ & $\alpha=1$
   \\\hline
  2& &0.343&0.358&0.384&0.559&0.804&1.177 \\
  5& &0.513&0.530&0.593&1.340&4.806&5.324 \\
  10& &0.760&0.817&0.945&10.471&11.389&12.197  \\
  20& &1.290&1.429&2.069&29.979&38.830&47.064  \\ \hline

 &&&&$\varepsilon=5\%$&&&\\
  \hline \vspace{1mm}
  N&& $\alpha=0$ & $\alpha=0.1$ & $\alpha=0.2$ & $\alpha=0.5$ & $\alpha=0.75$ & $\alpha=1$
   \\\hline
  2& &4.425&0.888&0.533&0.593&0.849&1.202 \\
  5& &18.816&1.077&0.662&1.565&4.985&5.437 \\
  10& &41.312&0.951&1.022&10.646&11.470&12.600  \\
  20& &145.172&1.517&2.273&30.339&39.561&47.072  \\
  \hline

  &&&&$\varepsilon=10\%$&&&\\
  \hline \vspace{1mm}
  N&& $\alpha=0$ & $\alpha=0.1$ & $\alpha=0.2$ & $\alpha=0.5$ & $\alpha=0.75$ & $\alpha=1$
   \\\hline
  2& &11.554&4.605&0.945&0.694&0.923&1.294 \\
  5& &43.446&4.075&0.749&1.758&5.052&5.449 \\
  10& &143.395&1.325&1.091&10.720&11.454&12.648  \\
  20& &503.319&1.648&2.422&30.776&39.758&47.693  \\
  \hline

  &&&&$\varepsilon=20\%$&&&\\
  \hline \vspace{1mm}
  N&& $\alpha=0$ & $\alpha=0.1$ & $\alpha=0.2$ & $\alpha=0.5$ & $\alpha=0.75$ & $\alpha=1$
   \\\hline
  2& &32.696&24.612&9.955&1.118&1.190&1.475 \\
  5& &132.542&53.841&1.869&2.171&5.362&5.625 \\
  10& &441.209&19.233&1.241&10.751&11.751&12.745  \\
  20& &1613.373&1.930&3.742&31.361&40.292&49.644  \\
 \hline
\end{tabular}
\caption{Simulation based estimates of the mean square error, when
$T=10N$ }\label{table2}
\end{table}

\begin{table}[ptb]
\begin{tabular}{cccccccc}
  \hline
  &&&&$\varepsilon=0\%$&&&\\
  \hline \vspace{1mm}
  N&& $\alpha=0$ & $\alpha=0.1$ & $\alpha=0.2$ & $\alpha=0.5$ & $\alpha=0.75$ & $\alpha=1$
   \\\hline
  2& & 0.035 & 0.035 & 0.039 & 0.051  & 0.067   & 0.084  \\
  5& & 0.050 & 0.052 & 0.057 & 0.087 & 0.135 & 0.204 \\
  10& & 0.075 & 0.081 & 0.093 & 0.185 & 0.395 & 8.703  \\
  20& & 0.129 & 0.142 & 0.181 &  0.910  & 28.127 & 31.790  \\
 \hline

 &&&&$\varepsilon=5\%$&&&\\
  \hline \vspace{1mm}
  N&& $\alpha=0$ & $\alpha=0.1$ & $\alpha=0.2$ & $\alpha=0.5$ & $\alpha=0.75$ & $\alpha=1$
   \\\hline
  2& & 2.504 & 0.304 & 0.076 & 0.060 & 0.068 & 0.092  \\
  5& & 10.863 & 0.207  & 0.066 & 0.092 & 0.136 & 0.217  \\
  10& & 37.549 & 0.129 & 0.100 & 0.191 &  0.409 & 9.329 \\
  20& & 136.364 & 0.157 & 0.194 &  0.891  & 28.354 & 33.200 \\
  \hline

  &&&&$\varepsilon=10\%$&&&\\
  \hline \vspace{1mm}
  N&& $\alpha=0$ & $\alpha=0.1$ & $\alpha=0.2$ & $\alpha=0.5$ & $\alpha=0.75$ & $\alpha=1$
   \\\hline
  2& & 8.910 & 2.493  & 0.285  & 0.066 & 0.073 & 0.096  \\
  5& & 39.015 & 2.203 & 0.089  & 0.098 & 0.142 & 0.240  \\
  10& & 133.655 & 0.404 & 0.107 & 0.207 & 0.474 & 9.746  \\
  20& & 493.386 & 0.182 & 0.203 & 1.249 & 28.467 & 33.925  \\
  \hline

  &&&&$\varepsilon=20\%$&&&\\
  \hline \vspace{1mm}
  N&& $\alpha=0$ & $\alpha=0.1$ & $\alpha=0.2$ & $\alpha=0.5$ & $\alpha=0.75$ & $\alpha=1$
   \\\hline
  2& & 28.470 & 18.606 & 7.524 & 0.106 & 0.102 & 0.115  \\
  5& & 124.702 & 44.374 & 0.327 & 0.113 & 0.168 & 0.272  \\
  10& & 429.087 & 17.016 & 0.128 & 0.232 & 0.592 & 10.423  \\
  20& & 1576.600  & 0.335  & 0.229   & 3.460 & 29.091 & 34.815 \\
 \hline
\end{tabular}
\caption{Simulation based estimates of the mean square error, when
$T=100N$}\label{table3}
\end{table}
\end{center}
We present simulation based estimates of the mean square error
given by
\begin{equation*}
\widehat{\mathrm{MSE}}=\frac{1}{n_{s}}\sum_{i=1}^{n_{s}}\|\widehat{\theta}_{i}-\theta_{0}\|^{2}
\end{equation*}
where $n_{s}$ is the number of samples (in our case $n_{s}=1000$),
$\theta_{0}=(\mu_{0}^{t},\mathrm{vech}(\Sigma_{0})^{t})^{t}$ and
$\widehat{\theta}_{i}=(\widehat{\mu}_{i}^{t},\mathrm{vech}(\widehat{\Sigma}_{i})^{t})^{t}$
is an estimation corresponding to the sample $i$. Here
$\mathrm{vech}(\Sigma)$ is ``the vector half'', namely the
$n(n+1)/2$-dimensional column vector obtained by stacking the
columns of the lower triangle of $\Sigma$, including the diagonal,
one below the other. Table \ref{table2} and Table \ref{table3}
present simulation based estimates of the mean square error, when
the sample size is $T=10N$, respectively when $T=100N$. When there
is no contamination, the MLE ($\alpha=0$) performs the best,
whatever the dimension $N$. On the other hand, the estimations
obtained with the minimum pseudodistance estimators in this case
are close to those provided by MLE, when $\alpha$ is not far to
zero (for example $\alpha=0.1$ and $\alpha=0.2$). In the presence
of contamination, the minimum pseudodistance estimators give much
better results than the MLE, in all considered cases. In most
cases, the choice $\alpha=0.2$ provides the best results in terms
of robustness. In the meantime, this choice corresponds to an
estimation procedure with high asymptotic relative efficiency,
according to the results from Table \ref{table1}. These facts
recommend $\alpha=0.2$ as a good choice in terms of trade-off
robustness efficiency. When the contamination is more pronounced,
i.e. $\varepsilon=10\%$ or $\varepsilon=20\%$, and the dimension
$N$ is low, i.e. $N=2$, the choices $\alpha=0.5$, $\alpha=0.75$ or
even $\alpha=1$ provide better robust estimates, but the
asymptotic relative efficiencies of the corresponding estimation
procedures are unacceptably low. Thus, values of $\alpha$ close to
zero, such as $\alpha=0.1$, $\alpha=0.2$, represent choices that
offer an equilibrium between robustness and efficiency. The
simulation results presented in Table \ref{table2} and Table
\ref{table3} shows that increasing sample size leads to improved
estimations.

\begin{figure}[]
\begin{center}
\includegraphics[width=9cm,height=9cm]{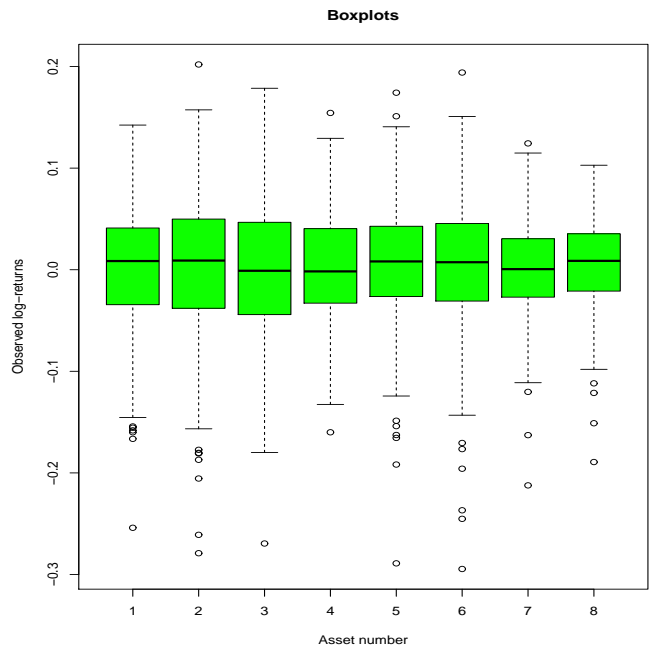}
\end{center}
\caption{Boxplots for monthly log-returns corresponding to the
MSCI Indexes (1: France, 2: Germany, 3: Italy, 4: Japan, 5:
Pacific ex JP, 6: Spain, 7: United Kingdom, 8: USA)
}\label{figure3}
\end{figure}

\section{Application for financial data}
We analyze 172 monthly log-returns of 8 MSCI Indexes (France,
Germany, Italy, Japan, Pacific Ex JP, Spain, United Kingdom and
USA) from January 1998 to April 2012 with the aim to construct
robust and efficient portfolios. The data are provided by MSCI
(www.msci.com). Boxplots for these data are presented in Figure
\ref{figure3}.

\begin{figure}
\begin{tabular}{ c  c }
 \includegraphics[width=7cm,height=7.5cm]{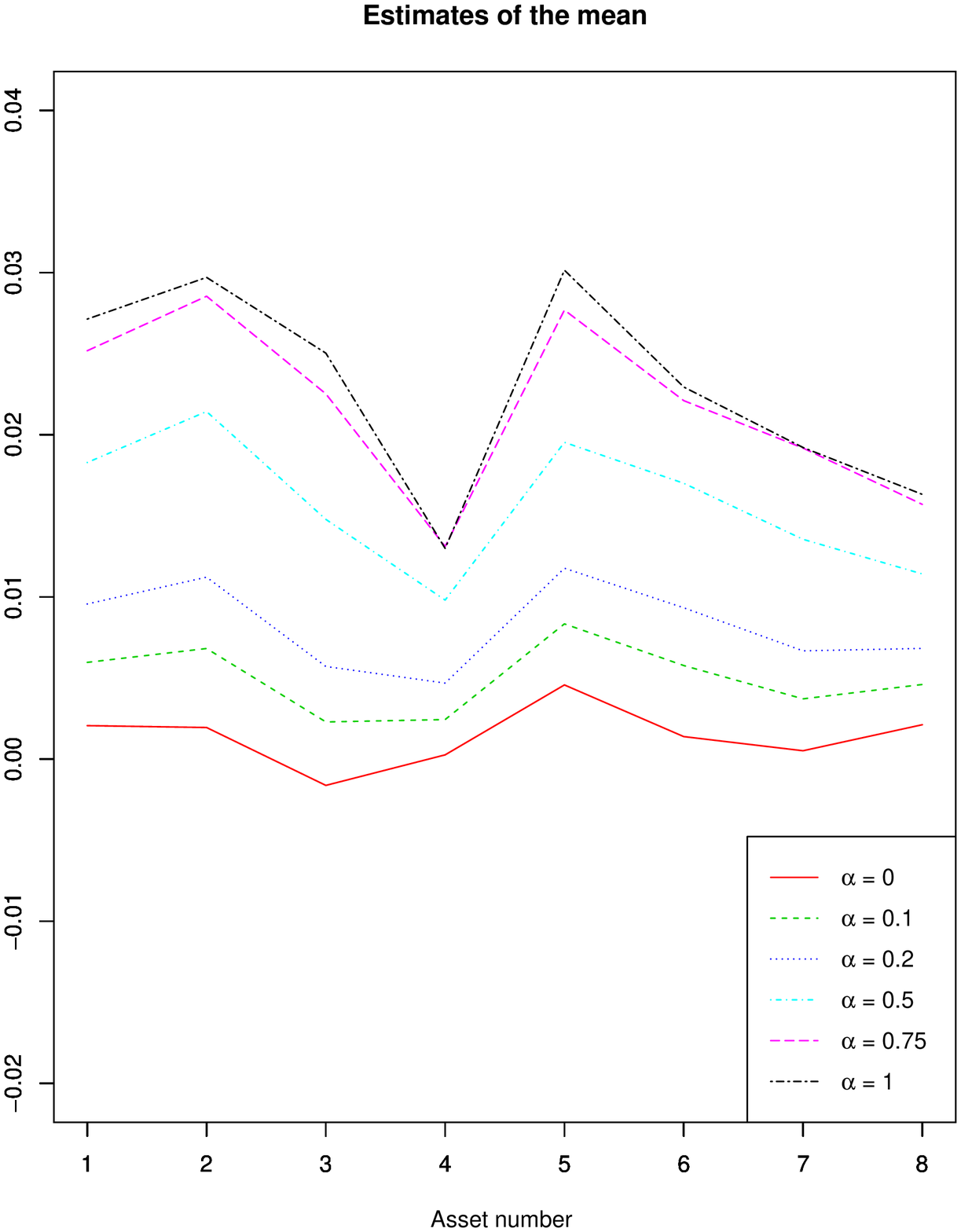}
 &   \includegraphics[width=7cm,height=7.5cm]{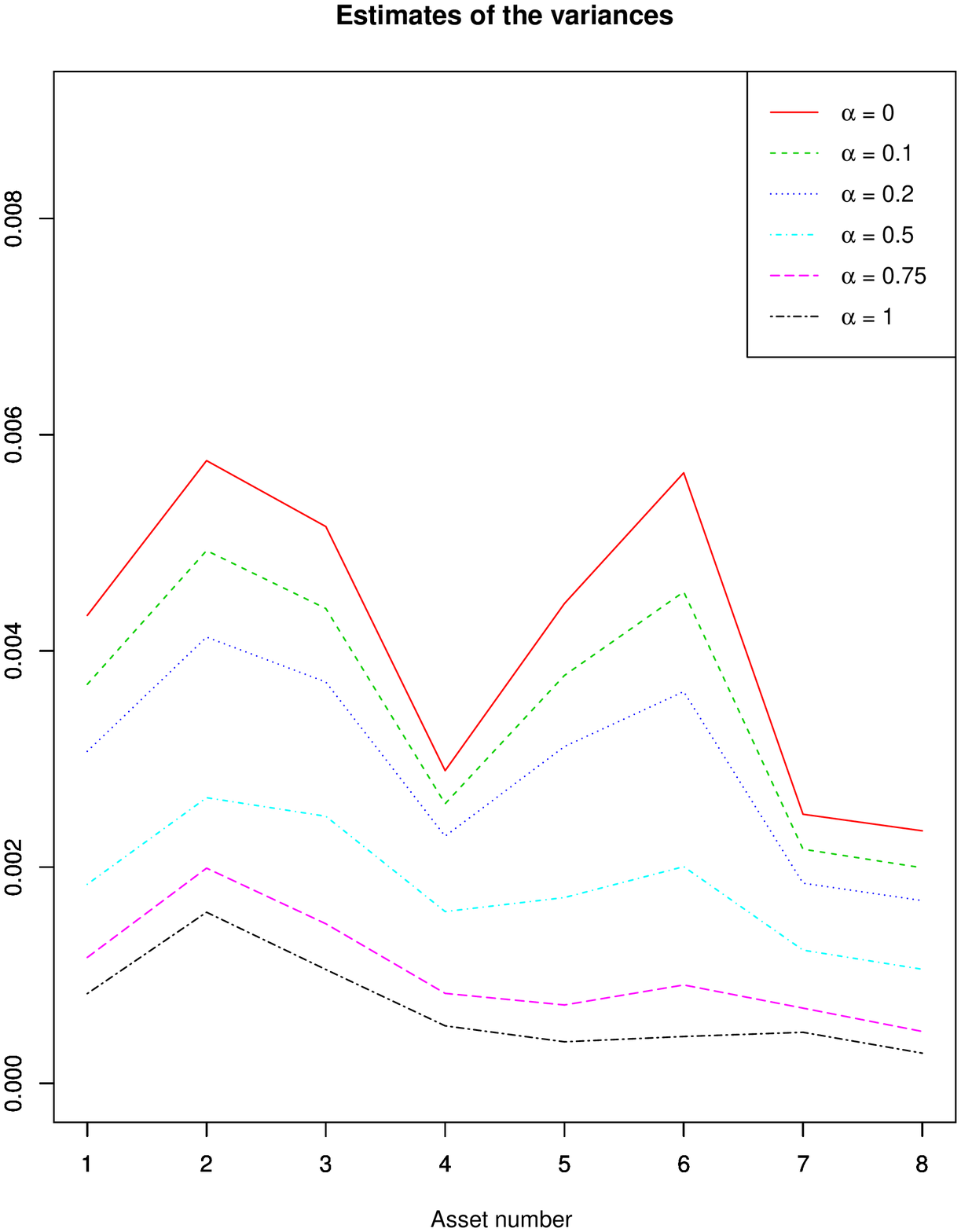}
\end{tabular}
\vspace{-2mm}\caption{Expected returns estimates (left) and
variance estimates (right) for the 8 MSCI Indexes (1: France, 2:
Germany, 3: Italy, 4: Japan, 5: Pacific ex JP, 6: Spain, 7: United
Kingdom, 8: USA) }\label{figure4}
\begin{tabular}{ c  c }
 \includegraphics[width=7cm,height=7.5cm]{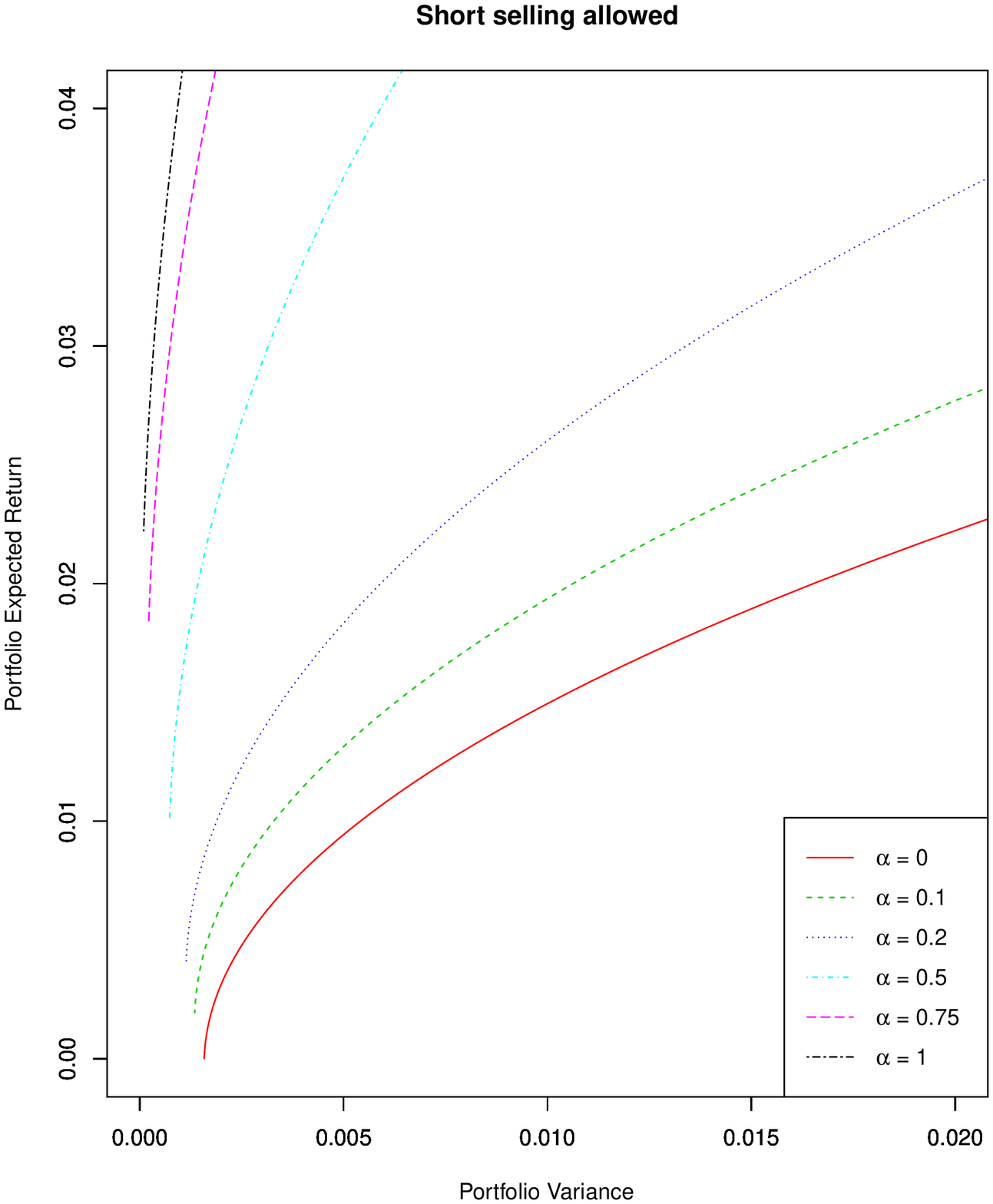}
 &   \includegraphics[width=7cm,height=7.5cm]{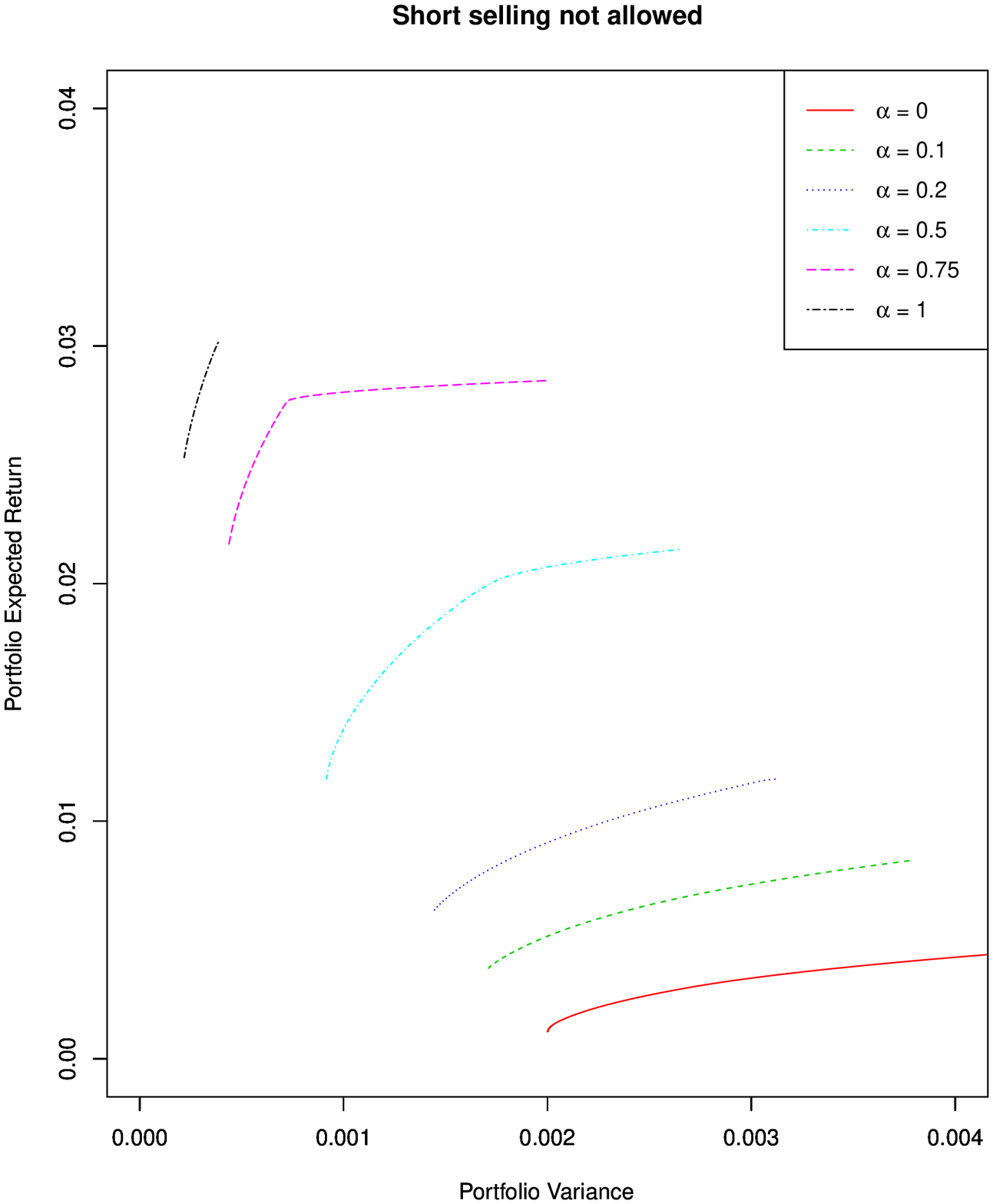}
\end{tabular}
\caption{Mean-variance efficient frontiers}\label{figure5}
\end{figure}

For these indexes, estimates of the expected return and of the
variance are represented in Figure \ref{figure4}. Note that the
estimates of the expected returns obtained with the minimum
pseudodistance estimators are larger than the maximum likelihood
ones. In the meantime, the minimum pseudodistance estimates of the
variances are smaller than those provided by the MLE.

Estimates of the mean vector and of the covariance matrix computed
with different minimum pseudodistance estimators are used to
determine efficient frontiers. In Figure \ref{figure5} we plot
efficient frontiers for the case ``short selling allowed'',
respectively for the case ``short selling not allowed''. In both
cases, the frontiers based on the minimum pseudodistance
estimations dominate those based on the classical maximum
likelihood estimations, yielding portfolios with larger expected
returns and smaller risks. Thus, the robust estimates reduce the
volatility effects which typically affects the results of the
traditional approaches.

The next step in our analysis is to identify the influential
observations which are responsible for the shift of the efficient
frontier. We perform this study in the case ``short selling''. In
this sense, we use the data influence measure (DIM) as diagnostic
tool (see Perret-Gentil and Victoria-Feser (2005)). This is
defined as the Euclidian norm of the influence function of the
estimator of weights based on maximum likelihood estimators of
$\mu$ and $\Sigma$. More precisely,
\begin{equation*}
\mathrm{DIM}(x,\widehat{p^{*}})=[\mathrm{IF}(x;p^{*},P_{\mu,\Sigma})^{t}\mathrm{IF}(x;p^{*},P_{\mu,\Sigma})]^{1/2}
\end{equation*}
where $\mathrm{IF}(x;p^{*},P_{\mu,\Sigma})$ is given by (\ref{f3})
with $\mathrm{IF}(x;\mu,P_{\mu,\Sigma})$ and
$\mathrm{IF}(x;\Sigma,P_{\mu,\Sigma})$ given by the formulas
(\ref{IF11}) and (\ref{IF22}) in the case $\alpha=0$. In order to
compute $\mathrm{DIM}$, the true parameters values
$\mu,\Sigma,p^{*}$ have to be known. In practice, these parameters
should be estimated in a robust way, such that DIM is not affected
by the outlying observations it is supposed to detect.

\begin{figure}
\begin{tabular}{ c  c }
\includegraphics[width=7.5cm,height=7.5cm]{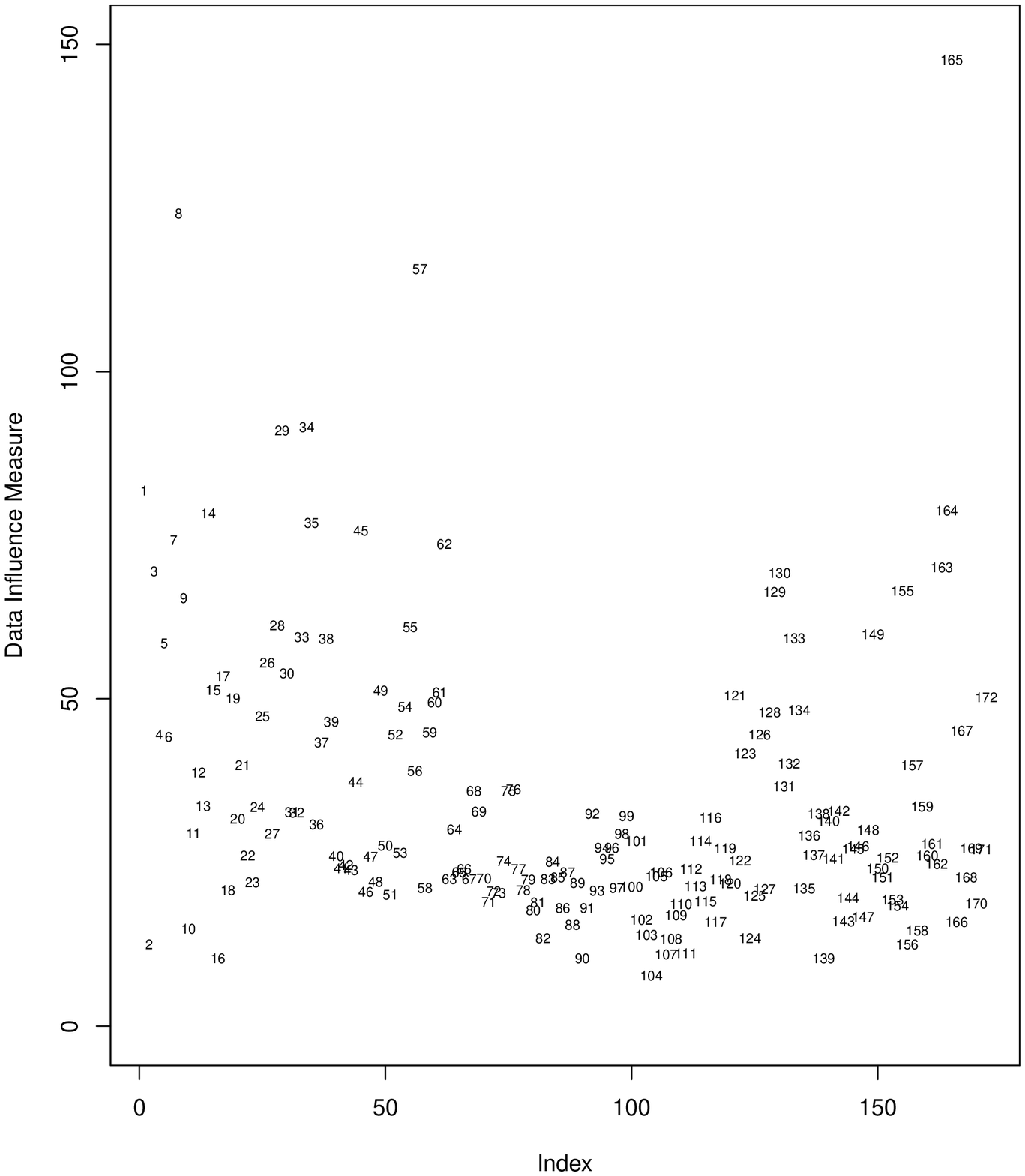} &
\includegraphics[width=7.5cm,height=7.5cm]{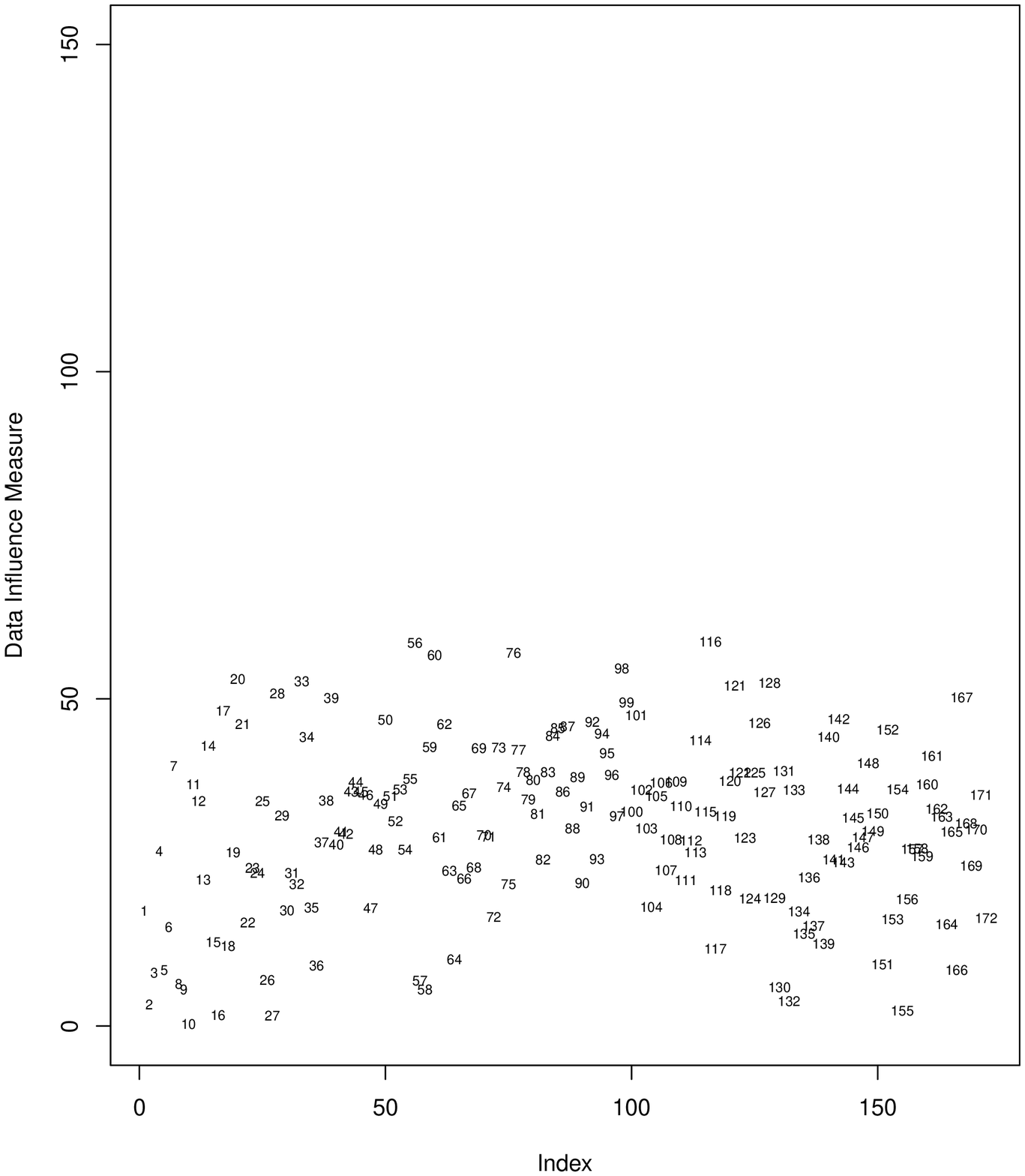}
 \end{tabular}
 \caption{The influence of each of the 172 observations on the classical estimator of
optimal portfolio weights (left hand side), respectively on the
robust estimator of optimal portfolio weights (right hand side),
as measured by DIM. The portfolio variance has been set to
0.005.}\label{figure6}
\end{figure}

In Figure \ref{figure6} (left hand side) we represent the
influence of each of the 172 observations on the estimator of the
optimal portfolio weights based on maximum likelihood estimators
of $\mu$ and $\Sigma$. Since $\mathrm{DIM}$ is related to a
specific portfolio on the efficient frontier, we made a choice,
namely the level of the portfolio variance has been set to 0.005.
The necessary robust estimates of $\mu,\Sigma,p^{*}$ have been
obtained with minimum pseudodistance estimators corresponding to
$\alpha=0.2$. The most influential observations as detected by DIM
correspond to negative economic events associated with known
financial crisis periods: 1998 Russian financial crisis (August
1998), ``dot-com crash'' of 2000-2002 and 2007-2012 global
financial crisis. On the other hand, the influence of these
observations is substantially reduced when using robust
procedures. This can be seen in the right hand side of Figure
\ref{figure6} where we represent the influence of each of the 172
observations on the robust estimator of the optimal portfolio
weights based on the minimum pseudodistance estimators of $\mu$
and $\Sigma$ corresponding to $\alpha=0.2$. Reducing the influence
of outlying observations leads to optimal portfolios with higher
returns and smaller variances.

Our theoretical and numerical results show that the optimal
portfolios based on minimum pseudodistance estimators are more
robust to extreme events than those obtained by plugging-in the
MLEs. When $\alpha$ is not far from 0, the minimum pseudodistance
estimators of $\mu$ and $\Sigma$ combine robustness with high
efficiency and these qualities are transferred to the portfolio
weights estimator. The numerical results based on simulations or
real data show that $\alpha=0.2$ represents a good choice in terms
of robustness and efficiency. All these recommend the new
procedure as a viable alternative to existing robust portfolio
selection methods.

\end{document}